\documentstyle[epsfig]{elsart}
\newcommand{\lsim}{\mathrel{\rlap{\lower4pt\hbox{\hskip0pt$\sim$}}
\raise1pt\hbox{$<$}}}
\newcommand{\gsim}{\mathrel{\rlap{\lower4pt\hbox{\hskip0pt$\sim$}}
\raise1pt\hbox{$>$}}}
\newcommand{\sfrac}[2]{\mbox{\footnotesize $\frac{#1}{#2}$}}
\begin{document}
\begin{frontmatter}
%_______________________ Title, Authors ____________________________________
% \hspace*{\fill}{Preprint Numbers: \parbox[t]{100mm}{ANL-PHY-8995-TH-98
%        \hspace*{\fill} nucl-th/97mmnnn\\
%        KSUCNR-103-97}}

\title{Infrared Behaviour of Propagators and Vertices}
\author[ua]{Frederick T. Hawes,}
\author[anl]{Pieter Maris}
\author[anl]{and Craig D. Roberts}
\address[ua]{Centre for the Subatomic Structure of Matter, University of
     Adelaide, \\Adelaide SA 5005, Australia}
     \address[anl]{Physics Division, Bldg. 203, Argonne National
     Laboratory,\\ Argonne IL 60439-4843, USA}
\begin{abstract}
We elucidate constraints imposed by confinement and dynamical chiral symmetry
breaking on the infrared behaviour of the dressed-quark and -gluon
propagators, and dressed-quark-gluon vertex.  In covariant gauges the
dressing of the gluon propagator is completely specified by ${\cal P}(k^2):=
1/[1+\Pi(k^2)]$, where $\Pi(k^2)$ is the vacuum polarisation.  In the absence
of particle-like singularities in the dressed-quark-gluon vertex, extant
proposals for the dressed-gluon propagator that manifest ${\cal P}(k^2=0)=0$
and ${\rm max}({\cal P}(k^2)) \sim 10$ neither confine quarks nor break
chiral symmetry dynamically.  This class includes all existing estimates of
${\cal P}(k^2)$ via numerical simulations.
\end{abstract}
\begin{keyword}
Gluon and quark Schwinger functions;
Dynamical Chiral Symmetry Breaking; Confinement;
Dyson-Schwinger equations; Lattice-QCD\\[2mm] 
\end{keyword}
\end{frontmatter}
Strong interaction phenomena are characterised by dynamical chiral symmetry
breaking (DCSB) and colour confinement.  At low energy, DCSB is the more
important; for example, in its absence the $\pi$- and $\rho$-mesons would be
nearly degenerate and at the simplest observational level that would lead to
a markedly different line of nuclear stability.  These phenomena can be
related to the infrared behaviour of elementary Schwinger functions in QCD
and herein we elucidate some constraints they place on this behaviour.

As described pedagogically in Ref.$\,$\cite{anu}, DCSB can be studied using the
QCD ``gap equation''; i.e., the Dyson-Schwinger equation (DSE) for the
renormalised dressed-quark propagator (connected, $2$-point, dressed-quark
Schwinger function), $S(p)$:
%
% \footnote{We employ a Euclidean space formulation with
% $\{\gamma_\mu,\gamma_\nu\}=2\delta_{\mu\nu}$, $\gamma_\mu^\dagger =
% \gamma_\mu$ and $a\cdot b=\sum_{i=1}^4 a_i b_i$.  A spacelike vector,
% $k_\mu$, has $k^2>0$.} 
%
\begin{eqnarray}
\label{gendse}
S(p)^{-1} & = & i\gamma\cdot p \,A(p^2) + B(p^2) \equiv 
        \frac{1}{Z(p^2)}\left[i\gamma\cdot p  + M(p^2) \right]\\
\label{gendseb}
& = & Z_2 (i\gamma\cdot p + m_{\rm bm})
+\, Z_1 \int^\Lambda_q \,
g^2 D_{\mu\nu}(p-q) \gamma_\mu S(q)
\Gamma_\nu(q,p) \,,
\end{eqnarray}
where $D_{\mu\nu}(k)$ is the renormalised dressed-gluon propagator,
$\Gamma_\mu(q;p)$ is the renormalised dressed-quark-gluon vertex, and
$\int^\Lambda_q := \int^\Lambda d^4 q/(2\pi)^4$ represents mnemonically a
translationally-invariant regularisation of the integral, with $\Lambda$ the
regularisation mass-scale.  The final stage in any calculation is to take the
limit $\Lambda\to\infty$.  In (\ref{gendse}), $Z_1$ and $Z_2$ are the
renormalisation constants for the quark-gluon vertex and quark wave function,
and $m_{\rm bm}$ is the current-quark bare mass: the chiral limit is obtained
with $m_{\rm bm}=0$$\,$\cite{mrt98,mr97}.

This equation is relevant because an order parameter for DCSB is the
chiral-limit, vacuum quark condensate$\,$\cite{mrt98}:
\begin{equation}
\langle \bar q q \rangle_\mu^0 :=
Z_4(\mu^2,\Lambda^2)\, N_c \int^\Lambda_q\,{\rm tr}_D
        \left[ S_{\hat m \propto m_{\rm bm} =0}(q) \right]\,,
\end{equation}
where  
$ Z_4(\mu^2,\Lambda^2) = [\alpha(\Lambda^2)/\alpha(\mu^2)]^{\gamma_m ( 1 +
\xi/3)}$
at one-loop order, with $\mu^2$ the renormalisation point, $\xi$ the
covariant-gauge fixing parameter [$\xi=0$ specifies Landau gauge] and
$\gamma_m= 12/(33-2 N_f)$ the gauge-independent mass anomalous dimension.
The $\xi$-dependence of $ Z_4(\mu^2,M^2) $ is just that required to ensure
that $\langle \bar q q \rangle_\mu^0$ is gauge independent.  It follows from
(\ref{gendse}) that an equivalent order parameter is
$
{\cal X}:= B(p^2=0)\,.
$
Chiral symmetry is dynamically broken when, with $m_{\rm bm}=0$, ${\cal
X}\neq 0$.

Equation$\,$(\ref{gendseb}) is a nonlinear integral equation and the properties
of its solution depend on the kernel, which is constructed from
$D_{\mu\nu}(k)$ and $\Gamma_\mu(q,p)$.  As summarised in Ref.$\,$\cite{mrp}, the
connected, dressed-gluon $2$-point function, $D_{\mu\nu}(k)$, satisfies an
oft analysed DSE.  The qualitative conclusion of these DSE studies is that if
the ghost-loop in the gluon DSE is unimportant, which is tautological in the
ghostless axial gauges, then relative to the free gauge boson propagator the
dressed-gluon propagator is significantly enhanced in the vicinity of
$k^2=0$, where it is a regularisation of $1/k^4$ as a
distribution$\,$\cite{bp89}.  That enhancement persists to $k^2\sim
1$-$2\,$GeV$^2$, where a perturbative analysis becomes quantitatively
reliable.

The other term in the kernel of (\ref{gendse}) is $\Gamma_\mu(q,p)$, the
connected, irreducible, renormalised dressed-quark-gluon vertex: $p$, $q$ are
the momentum labels of the quark and antiquark lines, and the total momentum
$P:=p-q$.  The analogue of this vertex in QED has been much studied and it is
argued$\,$\cite{bc80} that it should not exhibit particle-like singularities at
$P^2=0$.\footnote{
A particle-like singularity is one of the form $(P^2)^{-\alpha}$, $\alpha \in
(0,1]$. In this case one can write a spectral decomposition for the vertex in
which the spectral densities are non-negative.  This is impossible if
$\alpha>1$. $\alpha=1$ is the ideal case of an isolated, $\delta$-function
singularity in the spectral densities and hence an isolated, free-particle
pole.  $\alpha \in (0,1)$ corresponds to an accumulation, at the particle
pole, of branch points associated with multiparticle production.
} The reasoning is simple: such singularities do not arise at low order in
perturbation theory and hence such a vertex contradicts perturbation theory
in any domain on which a weak coupling expansion is valid.

Another arguably stronger reason is that a singularity of this type signals
the existence of a massless bound state that mixes with the gauge boson, and
such states have not been observed.  This feature can be elucidated by
considering the colour-singlet, axial-vector vertex in QCD, which is the
solution of
\begin{eqnarray}
\label{genave}
\left[\Gamma_{5\mu}^H(k;P)\right]_{tu} & = &
Z_2 \, \left[\gamma_5\gamma_\mu \frac{T^H}{2}\right]_{tu} \,+
\int^\Lambda_q \, [\chi_{5\mu}^H(q;P)]_{sr} \,K^{rs}_{tu}(q,k;P)\,,
\end{eqnarray}
where the matrix $T^{H}$ specifies the flavour structure of the vertex,
$\chi_{5\mu}^H(q;P) := {\cal S}(q_+) \Gamma_{5\mu}^H(q;P) {\cal S}(q_-)$ with
$q_+:= q+\eta_P P$, $q_-:= q-(1-\eta_P) P$ and $P$ the total momentum, and
${\cal S}(q) := {\rm diag}[S_u(q),S_d(q),S_s(q), \ldots]$.  In
(\ref{genave}), $K$ is the fully-amputated, quark-antiquark scattering
kernel: by definition it does not contain quark-antiquark to single
gauge-boson annihilation diagrams, such as would describe the leptonic decay
of the pion, nor diagrams that become disconnected by cutting one quark and
one antiquark line.

In the chiral limit the solution of this equation is$\,$\cite{mr97}
\begin{eqnarray}
\label{truavv}
\Gamma_{5 \mu}^H(k;P) & = &
\frac{T^H}{2} \gamma_5 
\left[ \rule{0mm}{5mm}\gamma_\mu F_R(k;P) + \gamma\cdot k k_\mu G_R(k;P) 
- \sigma_{\mu\nu} \,k_\nu\, H_R(k;P) \right]\\
&+ & \nonumber
 \tilde\Gamma_{5\mu}^{H}(k;P) 
+\,f_H\,  \frac{P_\mu}{P^2 } \,\Gamma_H^5(k;P)\,,
\end{eqnarray}
where: $F_R$, $G_R$, $H_R$ and $\tilde\Gamma_{5\mu}^{H}$ are regular as
$P^2\to 0$; $P_\mu \tilde\Gamma_{5\mu}^{H}(k;P) \sim {\rm O }(P^2)$; and
$\Gamma_H^5(k;P)$ is the Bethe-Salpeter amplitude for a massless pseudoscalar
bound state; i.e., $\Gamma_H^5(k;P)$ satisfies the associated homogeneous
Bethe-Salpeter equation.  The vertex is gauge covariant: the pole-position
and $f_H$, which is the leptonic decay constant, are gauge {\em
invariant}$\,$\cite{mrt98} and the bound state amplitude responds in a
well-defined manner to a gauge transformation.  In $3$-flavour, massless QCD
the poles in the axial-vector vertices correspond to the octet of Goldstone
bosons.  There should be no such singularities in the colour-singlet vector
vertex, and this is verified in model studies$\,$\cite{mf95}.

Similar observations apply to the fully-amputated, dressed-quark-gluon
vertex, $\Gamma_\mu(q,p)$.
% \footnote{The Lorentz structure of this vertex is such that 12 scalar form
% factors are required to specify it completely.  Not all the form factors
% are independent because of the Slavnov-Taylor identities, and only that
% multiplying $\gamma_\mu$ is ultraviolet divergent in perturbation theory.}
It satisfies an integral equation like (\ref{genave}) with the complication
that, in addition to the term involving $K$, there are $3$ other terms
involving the scattering kernels for: $q$-$\bar q$ to 2-gluon, $K^{2g}$;
$q$-$\bar q$ to ghost-antighost, $K^{gh\bar{gh}}$; and $q$-$\bar q$ to
3-gluon, $K^{3g}$.  Recall that, by definition, none of these kernels contain
single-gluon intermediate states.  Hence a massless, particle-like
singularity in this vertex signals the presence of a colour-octet bound state
in one of the scattering matrices: $M:= K/[ 1 - (SS)K]$; $M^{2g}:= K^{2g}/[ 1
- (DD)K^{2g}]$; etc.  As no such coloured bound states have been observed,
one must reject calculations or Ans\"atze for any of the Schwinger functions
that entail a particle-like singularity in this vertex.  The same objection
applies to particle-like singularities in the fully-amputated,
dressed-$3$-gluon vertex, and all like $n$-point functions.  This anticipates
the result of an estimate$\,$\cite{gluonv} of the $3$-gluon vertex via a
numerical simulation of lattice-QCD, which shows no evidence for a
singularity of any kind.

From a phenomenological perspective, a combination of $D_{\mu\nu}(k)$ enhanced
as described and $\Gamma_\mu(q,p)$ without particle-like singularities is an
excellent result, since it is {\em sufficient} to yield DCSB and
confinement\footnote{
Herein confinement means that the dressed-quark and -gluon $2$-point
functions do not have a Lehmann representation; i.e., do not have a spectral
representation with a non-negative spectral density.  This is a sufficient
but not necessary condition.\protect\cite{brs96}}
{\em without} fine-tuning$\,$\cite{anu}.  It can also provide for a
quantitatively accurate description of a wide range of hadronic
observables$\,$\cite{anu,mr97,pctrev}, although this depends more on the
detailed form of $D_{\mu\nu}(k)$ and $\Gamma_\mu(q,p)$.

Does the phenomenology of the strong interaction {\em require} that the gluon
propagator be strongly enhanced relative to the free gauge-boson propagator?
In Landau gauge
\begin{equation}
D_{\mu\nu}(k) = \left(\delta_{\mu\nu} - \frac{k_\mu
                k_\nu}{k^2}\right)\Delta(k^2) \,,\;
        \Delta(k^2):= \frac{1}{k^2}\,{\cal P}(k^2)\,,
\end{equation}
and the question can be posed as: Do the observable phenomena {\em
necessarily} require ($\Lambda^{N_f=4}_{\rm QCD} \sim 220\,$MeV)
\begin{equation}
\label{irenhanced}
{\cal P}(k^2) \gg 1\; \mbox{for} \; 0 < k^2 \lsim 10\,\Lambda_{\rm QCD}^2 \,?
\end{equation}  

We do not have an answer but we can explore alternatives.  The antithesis of
(\ref{irenhanced}) is the extreme\footnote{``Extreme'' because it corresponds
to a screening of the fermion-fermion interaction, as familiar in an
electrodynamical plasma, rather than the antiscreening often discussed in
zero-temperature chromodynamics.}  possibility that
\begin{equation}
\label{hype}
{\cal P}(k^2=0)=0\,,\;{\cal P}(k^2)\leq 1\;\forall\, k^2\,,
\end{equation}
which was canvassed in Ref.$\,$\cite{stingl}.  Therein the ghost-loop
contribution to the gluon DSE is retained and the Ans\"atze for the $3$-gluon
and quark-gluon vertices exhibit {\em ideal} particle-like poles [$\alpha =
1$].  Since these poles are an essential element of the solution procedure
then, in the absence of a physically sensible interpretation or explanation
of them, one could simply reject this result.

Alternatively, one can suppose that (\ref{hype}) is more robust than the
procedure employed to motivate it and explore the phenomenological
consequences of the conjecture$\,$\cite{stingl}: ${\cal P}_S(k^2):=
k^4/(k^4+b^4)$, where $b$ is a dynamically generated mass scale.  Following
this approach it was found that if there are no particle-like singularities
in the quark-gluon vertex, $\Gamma_\mu(q,p)$, then ${\cal P}_S(k^2)$ is
unable to confine quarks$\,$\cite{hawes,bender} and $b$ must be fine-tuned to
very small values [$b< b_c \simeq \Lambda_{\rm QCD}$] if DCSB is to
occur$\,$\cite{hawes,bender,natale}.   It is therefore apparent that (\ref{hype})
is phenomenologically difficult to maintain.\footnote{DCSB requiring $b \sim
0$ is indicative of the {\it dynamical evasion} of (\protect\ref{hype}) since
${\cal P}_S(k^2) \to 1$ rapidly for small values of $b$.  We do not consider
the possibility that an irreducible vertex has a non-particle-like
singularity; i.e., a singularity of the form $(k^2)^{-\alpha}$, $\alpha>1$,
as there is no indication of such behaviour in any study to date.}

Nevertheless, the hypothesis has been explored in studies$\,$\cite{marenzoni} of
the dressed-gluon $2$-point function using numerical simulations of
lattice-QCD.  ${\cal P}(k^2=0)$ is necessarily finite in simulations on a
finite lattice because of the inherent infrared cutoff.  Thus one can only
truly determine ${\cal P}(k^2 \sim 0)$ by considering the behaviour of the
numerical result in both the countable limit of infinitely many lattice sites
and the continuum limit.  The form ${\cal P}_S(k^2)$ does not provide as good
a fit to the lattice data as an alternative form, which in the countable
limit is
\begin{equation}
\label{Plat}
{\cal P}_L(k^2) := \frac{k^2}{M^2 + Z \,k^2
        \,\left( k^2 a^2 \right)^\eta}\,, \;
        0 < k^2 < 0.6/a^2 \, \sim 50 \,\Lambda_{\rm QCD}^2\,,
\end{equation}
where $1/a \approx 2.0\,$GeV is the inverse lattice spacing, $Z\approx
0.1$, $\eta \approx 0.53$, and $M\approx 0.16\,$GeV.  This takes the maximum
value ${\cal P}_L(k^2 = 21 \Lambda_{\rm QCD}^2) = 13.6$ and corresponds to a
less extreme alternative to (\ref{irenhanced}), which we shall characterise
as\footnote{A dressed-gluon propagator satisfying (\ref{hype}) automatically
satisfies (\ref{hypeB}).  The model of Ref.$\,$\cite{cornwall} is in the class
specified by (\ref{hypeB}), as are the fitted forms obtained in all existing
lattice-QCD simulations.}
\begin{equation}
\label{hypeB}
{\cal P}(k^2=0)=0\,,\;{\rm max}\left({\cal P}(k^2)\right) \lsim {\cal O}(10)\,.
\end{equation}
The feature ${\cal P}(k^2=0)=0$ is critically dependent on whether $M$ is
nonzero, or not.  It appears to be nonzero in the countable limit but, as
emphasised in Ref.$\,$\cite{marenzoni}, the behaviour of $M$ (and $\eta$) in the
continuum limit is unknown.

\begin{figure}[t]
\centering{\
\epsfig{figure=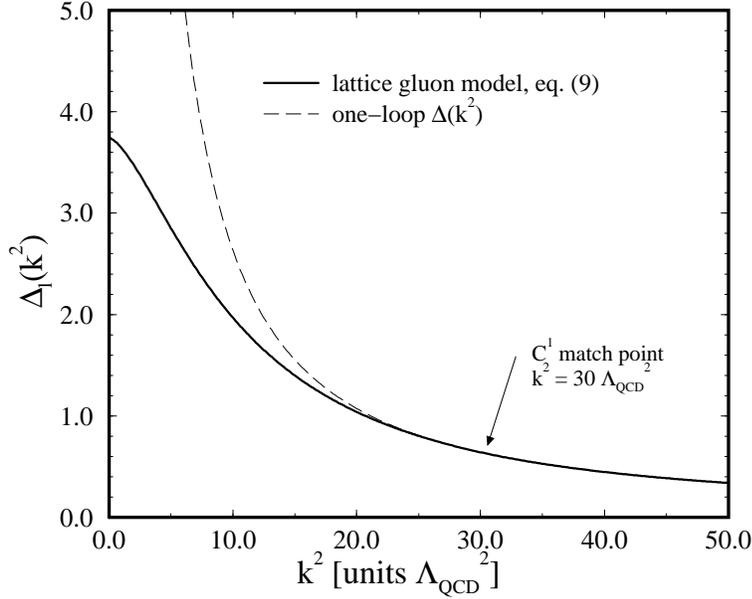,height=9.0cm}}
\caption{$\Delta_l(k^2):= {\cal P}_l(k^2)/k^2$ from
(\protect\ref{pourl}). ${\cal P}_l(k^2)$ is (\protect\ref{Plat}) in the
infrared and extrapolates this lattice model outside the domain accessible in
the simulation$\,$\protect\cite{marenzoni}.
\label{figgprop}}
\end{figure}

The phenomenological implications of (\ref{Plat}) can be explored using the
methods of Ref.$\,$\cite{hawes}.  A preliminary estimate follows by observing
that ${\cal P}_L(k^2)$ is approximately equivalent to ${\cal P}_S(k^2)$ if
one identifies $b_L \sim \sqrt{M/a} = 0.57\,$GeV.  Hence one expects
that (\ref{Plat}) does not generate DCSB nor confine quarks.\footnote{
A value of $b\approx 0.4\,$GeV$>b_c$ in ${\cal P}_S(k^2)$ provides the best
fit to the lattice data and this supports the same conclusion.
} In order to quantitatively verify this conclusion we note that: it is the
combination $g^2 {\cal P}(k^2)/k^2$ that appears in (\ref{gendse}) and $g^2$
is not determined in Ref.$\,$\cite{marenzoni}; and we must extrapolate ${\cal
P}_L(k^2)$ outside the fitted domain.  Both of these requirements are
fulfilled if we assume that a one-loop perturbative analysis is reliable for
\mbox{$k^2 \gsim 25 \,\Lambda_{\rm QCD}^2$}, and employ\footnote{
In representing ${\cal P}(k^2)$ in the ultraviolet via the running coupling,
$g^2(k^2)$, we are effectively enforcing the identity: $\tilde Z_1 = \tilde
Z_3$ between the ghost-gluon-vertex and ghost wave function renormalisation
constants.  This ``Abelian Approximation'' is a phenomenologically well
justified Ansatz that ensures$\,$\protect\cite{mr97} the correct one-loop
anomalous dimension for both: the dressed-quark propagator obtained as a
solution of (\protect\ref{gendse}); and the pseudoscalar meson bound state
amplitudes.  At most it introduces an error at large-$k^2$ in the power of
the $\ln$-dependence of ${\cal P}(k^2)$.
} 
\begin{equation}
\label{pourl}
g^2\,{\cal P}_l(k^2) := 
\left\{
\begin{array}{ll}
g_m^2 {\cal P}_L(k^2)\,,&  k^2\leq k_m^2\\
g^2(k^2)\,,               & k^2> k_m^2\,,
\end{array}
\right.
\end{equation}
requiring that $\Delta_l(k^2):= {\cal P}_l(k^2)/k^2$ and its first derivative
be continuous at $k_m^2$.  This procedure yields $\Delta_l(k^2)$ in
Fig.$\,$\ref{figgprop} with 
\begin{equation}
g_m = 0.65\,,\; k_m^2 = 30\,\Lambda_{\rm QCD}^2\,.
\end{equation}

It is now straightforward to solve (\ref{gendse}) with a variety of Ans\"atze
for the quark-gluon vertex that do not exhibit particle-like
singularities.\footnote{
Equation$\,$(\ref{pourl}) defines a renormalisable model quark DSE, which we
solved in the manner described in Ref.$\,$\protect\cite{mr97}.  For simplicity,
we renormalised at the momentum cutoff, $\Lambda_{\rm UV} \sim 10^4
\Lambda_{\rm QCD}$, since the $p^2$-evolution of $A(p^2)$ and $B(p^2)$ beyond
that point is completely determined by $g^2(k^2)$.
} We employed the bare vertex $\Gamma_\mu(p,q):= \gamma_\mu$; the
Ansatz$\,$\cite{bc80}:
\begin{eqnarray}
\label{bcvtx}
\lefteqn{i\Gamma^{\rm BC}_\mu(p,q):= i\Sigma_A(p^2,q^2)\,\gamma_\mu}\\ &&
\nonumber + (p+q)_\mu\,\left[\sfrac{1}{2}i\gamma\cdot (p+q) \,
\Delta_A(p^2,q^2) + \Delta_B(p^2,q^2)\right]\,,
\end{eqnarray}
where 
$\Sigma_A(p^2,q^2):= [A(p^2)+A(q^2)]/2$,
$\Delta_A(p^2,q^2):= [A(p^2)-A(q^2)]/[p^2-q^2]$ and
$\Delta_B(p^2,q^2):= [B(p^2)-B(q^2)]/[p^2-q^2]$;
and an augmented form$\,$\cite{cp90}
\begin{eqnarray}
\label{cpvtx}
\Gamma^{\rm CP}_\mu(p;q) & := & \Gamma^{\rm BC}_\mu(p,q) + 
        \Gamma^{6}_\mu(p,q)\,,\\
\Gamma^{6}_\mu(p,q) & :=  &
 \frac{\gamma_\mu (p^2 - q^2) - (p+q)_\mu (\gamma\cdot p - \gamma\cdot q)}
        {2 d(p,q)}\, \left[ A(p^2) - A(q^2) \right]
\,,
\end{eqnarray}
with $d(p,q):= ([p^2-q^2]^2 + [M(p^2)^2 + M(q^2)^2]^2)/(p^2+q^2)$, each of
which allows the quark DSE to be solved in isolation.  In all cases we found
${\cal X}=0$ in the chiral limit; i.e., no DCSB with $B(p^2)\equiv 0$.

\begin{figure}[t]
\centering{\
\epsfig{figure=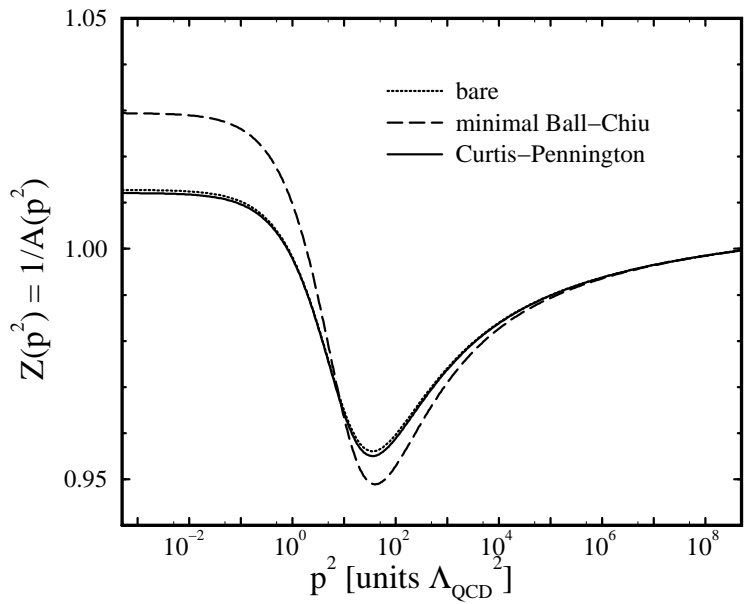,height=9.0cm}}
\caption{$Z(p^2)$ obtained as the solution of (\protect\ref{gendse}) using
(\protect\ref{pourl}) with: (\protect\ref{bcvtx}) - solid line;
(\protect\ref{cpvtx}) - dashed line; and $\Gamma_\mu(p;q)= \gamma_\mu$ -
dotted line.  That (\protect\ref{Plat}) does not confine quarks is manifest
in the result: $Z(p^2=0)\neq 0$, which is independent of the vertex Ansatz.
\label{figa}}
\end{figure}
The absence of DCSB means it is straightforward to decide whether
(\ref{pourl}) generates confinement.  In this case quark confinement is
manifest if $Z(p^2)$ is smooth and vanishes at $p^2=0$; while the existence
of a Lehmann representation and the concomitant lack of confinement is clear
if $Z(p^2)$ does not vanish at $p^2=0$.
%
% \footnote{The ultraviolet behaviour of $A(p^2)$ is fixed because our
% extrapolation of (\ref{Plat}) preserves the one-loop anomalous dimension of
% $S(p)$.}  
%
In Fig.$\,$\ref{figa} we plot the solution $Z(p^2)$ obtained from (\ref{gendse})
with the vertex Ans\"atze introduced above.  The behaviour of the solution is
qualitatively equivalent in each case.  This demonstrates explicitly that
(\ref{Plat}) neither generates DCSB nor confinement,\footnote{
The result of a recent numerical simulation$\,$\protect\cite{jis} is pointwise
smaller in magnitude $(\lsim 1/3)$ than (\protect\ref{Plat}) on the entire
fitted domain and hence the same conclusions apply in that case.  The
discrepancy is neither identified nor explained in Ref.$\,$\protect\cite{jis}.
Both simulations use $\beta = 6.0$.  Ref.$\,$\protect\cite{marenzoni} has $500$
configurations on a $24^3\times 48$ lattice with $\langle \partial_\mu
A_\mu(x)\rangle_{\rm Lat.}<10^{-6}$, while Ref.$\,$\protect\cite{jis} has $75$
configurations on a $32^3\times 64$ lattice with $\langle \partial_\mu
A_\mu(x)\rangle_{\rm Lat.}<10^{-12}$.
} thereby confirming our preliminary hypothesis based on the correspondence
with ${\cal P}_S$ via an effective value of $b$.

We have also solved the equation for $Z(p^2)$ using
\begin{equation}
\tilde{\cal P}_l(k^2):= \left( 1 
                + \varsigma\,{\rm e}^{-k^2/\Lambda_{\rm QCD}^2}\right)
                {\cal P}_l(k^2)
\end{equation}
where $\varsigma$ is a variable ``strength'' parameter.  Increasing
$\varsigma$ moves the peak in $\tilde{\cal P}_l(k^2)$ toward $k^2=0$ and
increases its height, thereby making it increasingly like the model of
Ref.$\,$\cite{mr97}.  The form of $Z(p^2)$ is qualitatively unchanged and hence
there is no signal for the onset of confinement until $\varsigma\gsim 300$.
At $\varsigma= 300$ the maximum value is
\begin{equation}
\tilde{\cal P}_l(k^2=0.98\,\Lambda_{\rm QCD}^2)= 210
\end{equation}
and 
$\tilde{\cal P}_l(0.98\,\Lambda_{\rm QCD}^2)/
        \tilde{\cal P}_l(30\,\Lambda_{\rm QCD}^2)= 16$, 
cf. 
${\cal P}_l(21\,\Lambda_{\rm QCD}^2)/
        {\cal P}_l(30\,\Lambda_{\rm QCD}^2)= 1.0$.
In the model of Ref.$\,$\cite{mr97} the peak is at $k^2= 3.7\,\Lambda^2_{\rm
QCD}$ and the value of this ratio is $44$, neglecting only for the purpose of
this comparison the purely long-range, $\delta^4(k)$-part of that
interaction.  A comparison of $g_m^2 \tilde{\cal P}_l(\Lambda_{\rm
QCD}^2)\approx 89$ with the critical coupling of $g_c^2 \approx 11$ in
Refs.$\,$\cite{atkinson} shows that such large values of $\varsigma$ ensure
DCSB.

The hypothesis (\ref{hype}) has also re-emerged recently in a DSE
study$\,$\cite{bloch} that is qualitatively akin to Ref.$\,$\cite{stingl}: in
its result that ${\cal P}(k^2)= A \,k^4$ for $k^2 \simeq 0$; in postulating a
significant role for the ghost-loop in the gluon propagator; and in employing
a ghost-gluon vertex that is free of particle-like singularities.\footnote{
A kindred study, Ref.$\,$\protect\cite{hauck}, employs Ans\"atze for the
ghost-gluon and $3$-gluon vertices that exhibit particle-like singularities.
Recall that no evidence for such behaviour is observed in lattice-QCD
estimates of the $3$-gluon vertex$\,$\protect\cite{gluonv}.
} Following the above analysis, the results of Refs.$\,$\cite{hawes,bender}
can be applied directly in this case.  The gluon propagator is smooth and can
be characterised by a value of $b^4 \approx \lambda^4/A$, where $A\sim 1$ and
$\lambda$ is a mass-scale that is left undetermined in Ref.$\,$\cite{bloch}.
Choosing any reasonable value of $\lambda$; e.g., $\lambda \gsim \Lambda_{\rm
QCD}$, this gluon propagator, with a quark-gluon vertex that is free of
particle-like singularities, neither yields DCSB nor confines quarks.

It was recognised in Ref.$\,$\cite{marenzoni} that the hypotheses (\ref{hype})
and (\ref{hypeB}) are problematic, and this is emphasised by our results: in
the absence of particle-like singularities in the dressed-quark-gluon vertex,
the possibility of a dressed-gluon propagator satisfying (\ref{hypeB}) is
excluded by the existence of DCSB.  In the lattice simulations the infrared
scale, $M$, that entails (\ref{hypeB}) may vanish in the continuum limit:
DCSB and confinement appear to require that.  However, this limit is
presently unexplored.  In the DSE study the treatment of the ghost-gluon
vertex and ghost-gluon scattering kernel is rudimentary and, with the
information currently available to us, it is difficult to estimate the
sensitivity of the results to the truncations.  The incompatibility we
identify between (\ref{hype}) and the phenomena of the strong interaction
suggest that an examination of the effects of these truncations is necessary.

Our primary predicate is that the dressed-quark-gluon vertex should not
exhibit a particle-like singularity.  If this is false and the vertex
exhibits a singularity at $P^2=0$ then (\ref{hypeB}), or even (\ref{hype}),
may be reconcilable with the phenomena of the strong interaction.  However,
whether that is truly the case will likely depend on details and therefore
require {\em fine-tuning} in the theory.

We acknowledge useful conversations with M. R. Pennington.  PM and CDR are
grateful to the staff of the Special Research Centre for the Subatomic
Structure of Matter at the University of Adelaide for their hospitality and
support during the Workshop on Nonperturbative Methods in Quantum Field
Theory, during which this work was conceived.  This work was supported in
part by the US Department of Energy, Nuclear Physics Division, under contract
no. W-31-109-ENG-38, and benefited from the resources of the National Energy
Research Scientific Computing Center.

%______________________________ References ______________________________

%%----------------------------------------
\end{document}